# Complete mapping of magnetic anisotropy for prototype Ising van der Waals FePS$_3$

Muhammad Nauman[†1], Do Hoon Kiem[†2], Sungmin Lee[3], Suhan Son[3,4], Je-Geun Park[3,4], Woun Kang[5], Myung Joon Han[*2], and Younjung Jo[*1]

[1]Department of Physics, Kyungpook National University, Daegu 41566, Korea.

[2]Department of Physics, Korea Advanced Institute of Science and Technology (KAIST), Daejeon 34141, Korea.

[3]Department of Physics and Astronomy, Seoul National University, Seoul 08826, Korea.

[4]Center for Quantum Materials, Seoul National University, Seoul 08826, Korea.

[5]Department of Physics, Ewha Womans University, Seoul 03760, Korea.

[†]These two authors contributed equally to this work

E-mail: mj.han@kaist.ac.kr, jophy@knu.ac.kr

## ABSTRACT

Several Ising-type magnetic van der Waals (vdW) materials exhibit stable magnetic ground states. Despite these clear experimental demonstrations, a complete theoretical and microscopic understanding of their magnetic anisotropy is still lacking. In particular, the validity limit of identifying their one-dimensional (1-D) Ising nature has remained uninvestigated in a quantitative way. Here we performed the complete mapping of magnetic anisotropy for a prototypical Ising vdW magnet FePS$_3$ for the first time. Combining torque magnetometry measurements with their magnetostatic model analysis and the relativistic density functional total energy calculations, we successfully constructed the three-dimensional (3-D) mappings of the magnetic anisotropy in terms of magnetic torque and energy. The results not only quantitatively confirm that the easy axis is perpendicular to the *ab* plane, but also reveal the anisotropies within the *ab*, *ac*, and *bc* planes. Our approach can be applied to the detailed quantitative study of magnetism in vdW materials.

## 1. Introduction

Two-dimensional (2-D) van der Waals (vdW) materials, including those with atomic-layer thickness, provide an exciting platform for the study of 2-D magnetism [1–10]. These 2-D magnets are increasingly receiving attention owing to their intriguing magnetic properties and potential for application as intercalation materials [11, 12] and in spintronic devices [13, 14]. Among the focuses in this line of study is magnetic anisotropy, which offers the possibility of stabilizing spin ordering under circumstances that are close to the ideal 2- D limit [6, 15–17]. Therefore, detailed information on magnetic anisotropy and its energy scale has been crucial in unveiling the fundamental nature of magnetism and in the vdW material based spintronics.

Among the many magnetic vdW materials, the layered transition metal trichalcogenide family, namely, $TM$PS$_3$ ($TM$ = Fe, Co, Mn, V, Zn, or Ni), provides a unique platform for conducting fundamental studies. In these antiferromagnetic (AFM) materials, whose magnetic and crystallographic structures are both 2-D [1, 5, 18–23], the vdW gap between the 2-D layer facilitates an indirect exchange interaction along the $c$-axis, resulting in a highly anisotropic magnetic behavior. Given that the anisotropy of these materials is closely related to the crystal field effect, orbital occupation, and symmetry [24], each of these compounds can serve as a material platform for the exploration of theoretical spin models in the 2-D limit [25]. For example, the magnetization axes in FePS$_3$ and MnPS$_3$ lie perpendicular to the layer planes, while in NiPS$_3$ they lie within the layer plane in the ordered state. Further, it is believed that FePS$_3$, NiPS$_3$, and MnPS$_3$ effectively represent the Ising-, XY- (or XXZ-), and Heisenberg-type spin models, respectively [22, 26–28]. Detailed investigations of their paramagnetic (PM) regimes have shown that the magnetic susceptibilities of MnPS$_3$, NiPS$_3$, and FePS$_3$ are isotropic, weakly anisotropic, and highly anisotropic, respectively [24].

Our current study is focused on FePS$_3$, which is unreactive in air; thus, is suitable for studying the hitherto unknown phenomena regarding 2-D magnetism and for realizing antiferromagnet-based spintronic functionalities [29-31]. However, the important details regarding its anisotropy profile and the energetic validity limit of its Ising nature are still unclear; i.e., the magnetic anisotropy of unstacked monocrystalline FePS$_3$ has not yet been reported, and its quantitative Ising nature is still largely unexplored. Particularly, the energy cost of the spin rotation from its easy axis to its hard axis alignment is still unknown. Additionally, the in-plane magnetic anisotropy energy, which must be zero in an ideal Ising system, but is likely to be non-zero in practice, has not yet been addressed, even though such a quantitative understanding represents a crucial step in fundamental research and application. Therefore, detailed measurements and theoretical calculations regarding the trichalcogenide family as well as other 2-D vdW systems are highly needed.

To investigate the Ising magnetic anisotropy and its validity limit, angle dependent torque magnetometer experiments were carried out using high-quality single crystals. Systematic theoretical analysis and total energy calculation were performed based on the magnetostatic model construction and the relativistic density functional theory (DFT) calculation. We successfully constructed the complete mapping of magnetic anisotropy for this prototype Ising vdW FePS$_3$ for the first time. It is expected that the results of this study will pave the way for a better understanding of the quantitative aspects of similar systems.

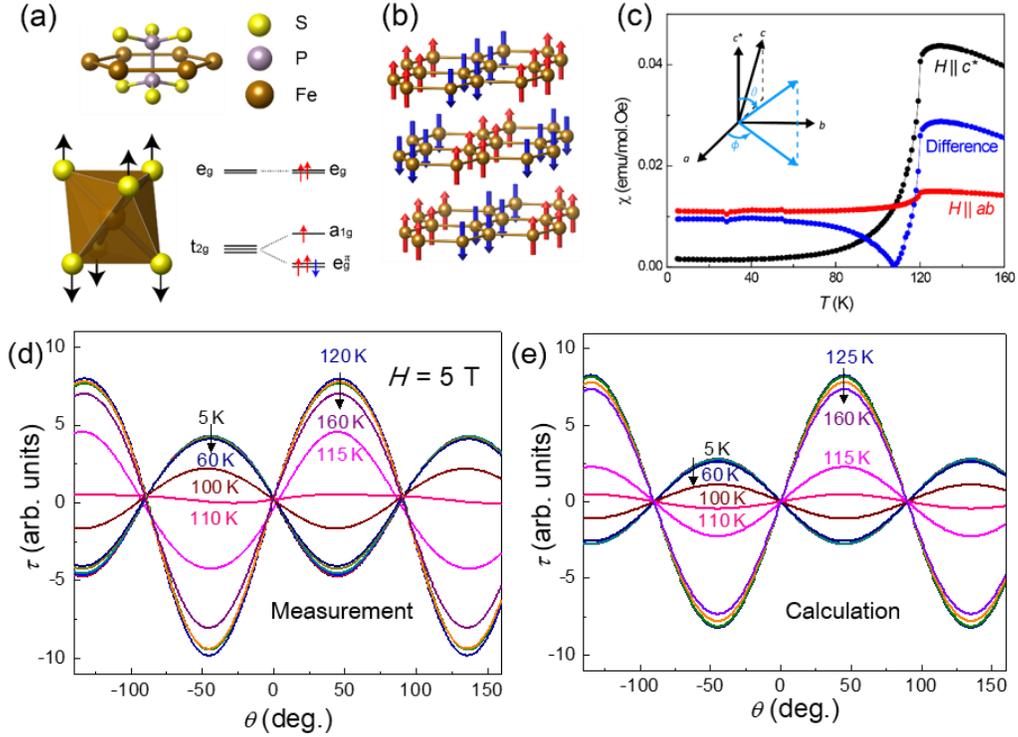

**Figure 1** (a) Local atomic structure around Fe atoms and the spin configuration. The yellow, bright violet, and brown spheres represent the S, P, and Fe atoms, respectively. The trigonal distortion of the $S_6$ octahedra around the Fe atoms is depicted by black arrows. Honeycomb layers of $[P_2S_6]^{4-}$ and $Fe^{2+}$ alternate along the *c*-axis. (b) The ground-state spin order of $FePS_3$. (c) The measured magnetic susceptibility as a function of temperature. The black, red, and blue curves correspond to the out-of-plane susceptibility, in-plane susceptibility, and their absolute difference, respectively. The inset shows the definitions of axes and field angle rotations. The *c*-axis is inclined at 108.37° with respect to the *a*-axis, and normal to the *b*-axis. *c\**- is perpendicular to the *ab* plane. (d) The experimental torque magnetometer data for various temperatures at *H* = 5 T. The field rotates out-of-plane. (e) The calculated torque, $\tau_{ac}(\theta)$, by using Eq. (2).

## 2. Results and discussion

### 2.1 Crystal structure, electronic configuration, and anisotropic antiferromagnetism in $FePS_3$

$FePS_3$ is an AFM Mott insulator with an optical gap of ~1.5 eV [32–34]. Its honeycomb lattice structure consists of $Fe^{2+}$ [$d^6$] ions, each of which is surrounded by six sulfur atoms with trigonal symmetry as shown in Fig. 1(a). The sulfur atoms are bonded to two phosphorus atoms, forming a dumbbell structure, and $[P_2S_6]^{4-}$ provides the local octahedral environment for $Fe^{2+}$. The elongated octahedron induces the splitting of $t_{2g}$ orbitals into $e_g^\pi$ and $a_{1g}$, and the high spin configuration is stabilized. $FePS_3$ has the 'zigzag-stripe' AFM spin ordering; see Fig. 1(b) [21, 24, 26, 35–37]. This AFM ground state is more stable than ferromagnetic phase by 31.7 meV/$Fe^{2+}$ (~ 127 T/$Fe^{2+}$). It has attracted significant attention owing to the presence of the same order down to the monolayer limit [26, 37]. Neutron single-crystal diffraction measurements have shown that the moment of Fe atom is ferromagnetically coupled with those of the two atoms that are closest to it and antiferromagnetically coupled with that of the third. Also, the previous neutron scattering and Mössbauer spectroscopy showed the Ising nature of the spins in this material [21, 38, 39] and the magnon data fitting on to the spin model found the large magnetic anisotropy [40].

## *2.2 3-D mapping of anisotropic torques and theoretical interpretation*

Fig. 1(c) shows our magnetic susceptibility data of FePS$_3$ as a function of temperature. The black and red dots represent the measurements along the out-of-plane and the in-plane directions, respectively. The PM-to-AFM phase transition is marked by the suppression of the susceptibility signals at $T_N$ = 118 K. Particularly, the suppression of the out-of-plane susceptibility ($\chi_\perp$ or $\chi_c$) is significantly more pronounced than that of the in-plane susceptibility ($\chi_\parallel$ or $\chi_{ab}$). As discussed in previous studies [24, 26], together with data from Mössbauer, and neutron scattering [21, 38, 39], this behavior can be regarded as key evidence for Ising nature of the spins in this material, which has an easy axis along the $c^*$-direction (defined as perpendicular to the *ab* plane; see inset of Fig. 1(c)). Given that the $\chi_\perp$ suppression is stronger than the $\chi_\parallel$ suppression, this eventually results in the magnitude reversal of these susceptibility components, i.e., $\chi_\perp < \chi_\parallel$ at $T \leq 108$ K. As discussed below, it generates the intriguing behaviors in our torque measurements.

Important details regarding magnetic anisotropy are revealed via angle-dependent torque measurements. Fig. 1(d) shows the torque, $\tau(\theta)$, that was measured by changing the azimuthal field angle relative to the $c^*$-axis (see the inset of Fig. 1(c)). In this measurement, the applied magnetic field of $H$ = 5 T was lying within the *ac* plane. At temperatures below $T_N$, $\tau(\theta)$ exhibits such behavior that its amplitude gradually decreases with increasing temperature. Remarkably, the sign of the amplitude of $\tau(\theta)$ is suddenly reversed when the temperature changes from 110 K to 120 K, and at $T$ = 120 K the amplitude reaches to the maximum. Note that this sudden change observed across $T_N$ is indicative of the response of the mechanical torque to the magnetic transition.

To understand this behavior as well as its physical implications, a macroscopic magnetostatic model was constructed. We note that, in the absence of an external field ($H$ = 0), the net magnetic moment is zero in both AFM and PM phases. However, there is a key difference between these two phases. In the PM phase ($T > T_N$), spins have the preferred direction along the $c^*$-axis while they are not ordered. Thus, the external field induces a moment along the easy axis direction ($c^*$-axis; $\theta = 0°$) by aligning the originally disordered spins. On the other hand, in the AFM phase ($T < T_N$), the up and down spins are paired with each other, resulting in a zero net moment. The field induces the moment most likely when it is perpendicular to the easy axis (i.e., $\theta = 90°$) given that the $H$-field strength is much smaller than the AFM pairing strength. Namely, the net moment in this case is induced mainly via spin tilting. This difference in the direction of the induced moment between the PM and AFM phases can be responsible for the 90° phase shift observed in Fig. 1(d). Taking this picture as our working hypothesis, we express the torque induced by the external magnetic field, $\vec{H}$, as follow:

$$\vec{\tau} = (\overleftrightarrow{\chi} \cdot \vec{H}) \times \vec{H}, \tag{1}$$

where $\vec{\tau}$ represents the torque per unit volume on a system, and $\overleftrightarrow{\chi}$ represents the magnetic susceptibility tensor. For the moment, we assume that $\overleftrightarrow{\chi}$ has zero off-diagonal components owing to the crystal symmetry, with only diagonal components contributing to the torque. We will see that this assumption is useful for understanding most of our experiments. Later, however, this assumption will be refined for understanding the detailed in-plane anisotropy (see below). Now, the directional torques read:

$$\tau_{ac}(\theta) = \tfrac{1}{2}(\chi_{cc} - \chi_{aa})H^2 \sin 2\theta \quad \text{for } \phi = 0° \quad (2)$$

$$\tau_{bc}(\theta) = \tfrac{1}{2}(\chi_{bb} - \chi_{cc})H^2 \sin 2\theta \quad \text{for } \phi = 90° \quad (3)$$

$$\tau_{ab}(\phi) = \tfrac{1}{2}(\chi_{aa} - \chi_{bb})H^2 \sin 2\phi \quad \text{for } \theta = 90° \quad (4)$$

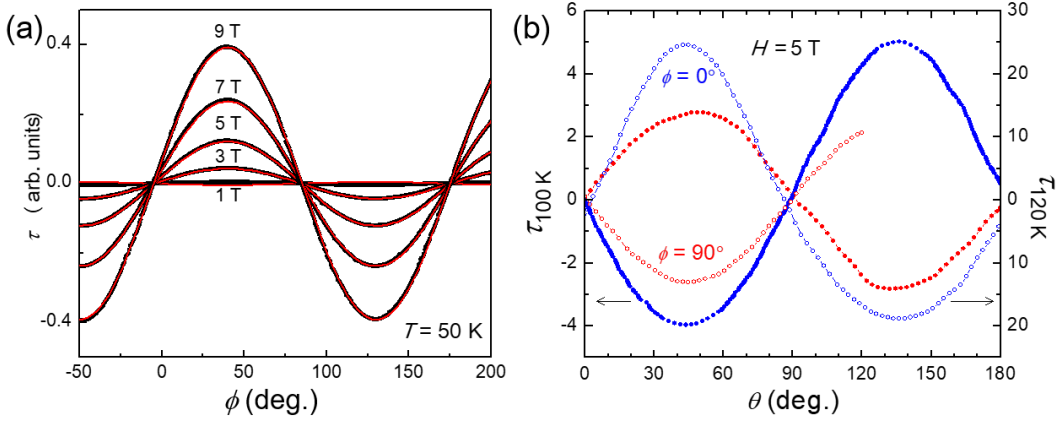

**Figure 2** (a) The measured torque, $\tau_{ab}(\phi)$ (black) along with the fitting lines (red) as functions of the in-plane field rotation angle, $\phi$, and the field strength. The temperature was set at 50 K. (b) The out-of-plane torque at temperatures below (solid symbols; left-hand Y-axis; $T$ = 100 K) and above (open symbols; right-hand Y-axis; $T$ = 120 K) $T_N$. The blue and red curves correspond to the measured torque in the field $H$ = 5 T on the $ac$ ($\phi$=0°) and $bc$ ($\phi$=90°) planes, respectively.

where $\tau(\theta)$ and $\tau(\phi)$ represent the torque as measured using the rotation of the magnetic field in the in-plane and out-of-plane field directions, respectively. Note that we distinguish two out-of-plane field directions explicitly. $\chi_{aa}$, $\chi_{bb}$, and $\chi_{cc}$ represent each directional component of the magnetic susceptibility tensor, and $\chi_{aa,bb}$ and $\chi_{cc}$ corresponds to $\chi_\parallel$ and $\chi_\perp$, respectively. Using the measured susceptibility as inputs (i.e., $\chi_\parallel$ and $\chi_\perp$ as shown in Fig. 1(c)), it is straightforward to obtain $\tau(\phi)$ and $\tau(\theta)$ from Eq. (2)–(4). Fig. 1(e) shows the calculation result of the out-of-plane torque, $\tau_{ac}(\theta)$, which is in excellent agreement with Fig. 1(d). This result confirms the relevance of our model and simultaneously reveals the detailed magnetic behavior of this material along the out-of-plane field direction. In particular, one can conclude that the sudden phase change of $\tau(\theta)$ across $T = T_N$ is attributed to a 90° change in the induced-moment direction.

Fig. 2(a) shows the in-plane anisotropy revealed by $\tau_{ab}(\phi)$, i.e., the torque measured via the variation of the in-plane magnetic field at $T = 50$ K $< T_N$. Emphatically, what we measured here shows the (Ising) spin anisotropy profile in the plane perpendicular to the spin direction. Compared with the out-of-plane torque, $\tau_{ac}(\theta)$, shown in Fig. 1(d), $\tau_{ab}(\phi)$ is order-of-magnitude smaller as expected. Importantly, however, it is still clearly non-zero. Once again, the overall behavior of $\tau_{ab}(\phi)$ can be well understood based on our model. The red lines in Fig. 2(a) represent the curves obtained from Eq. (4), which are in good agreement with the measured data shown in black. The data also show that, for $0° \leq \phi \leq 90°$, $\tau(\phi)$ is positive and $\chi_{aa} > \chi_{bb}$. Given that the torque is exerted to minimize the magnetostatic energy, the results show that the induced moment favors the alignment along the $b$-axis rather than the $a$-axis within the plane. This point will be quantitatively confirmed by our DFT total energy calculations (see below).

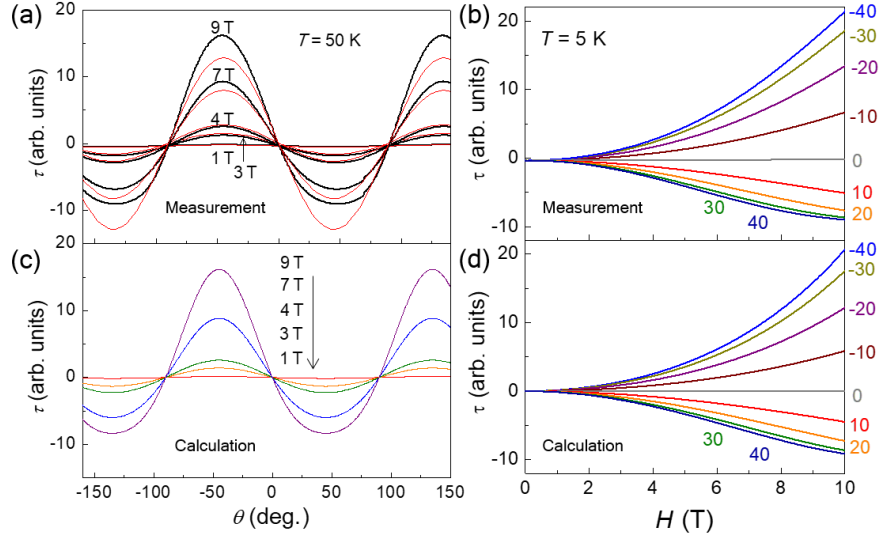

**Figure 3** (a) The out-of-plane torque, $\tau_{ac}(\theta)$ (black) measured with the rotating field in the ac plane at various field strengths and at a fixed temperature, $T = 50$ K. The red lines represent the fitting from the theoretical model of Eq. (2). (b) The field-dependent torque measured at various field orientations. The numbers show the field angles $\theta$ in the *ac* plane as defined by the inset of Fig. 1(c). Temperature was fixed at $T = 5$ K. (c) The calculated torque, $\tau_{ac}(\theta)$, by the updated formula of Eq. (6). (d) The calculated field-dependent torque based on Eq. (6). The off-diagonal components were fitted by $\chi_{ca}^0 = 0.2 \times \chi_{aa}, \chi_{ac}^0 = 0.4 \times \chi_{aa}$.

In Fig. 2(b), $\tau_{ac}(\theta)$ (i.e., the field rotation within the *ac* plane; $\phi = 0°$) is compared with $\tau_{bc}(\theta)$ (i.e., the field rotation within the *bc* plane; $\phi = 90°$). Here, the blue (red) color represents the data corresponding to $\phi = 0°$ (90°), and the filled (open) symbols refer to the results of $T = 100$ K (120 K). In $\tau_{ac}(\theta)$ and $\tau_{bc}(\theta)$, the sign changes are clearly observed below and above $T = T_N$ (i.e., comparing the filled vs. open symbol data), indicating the tilting of the effective moment by 90° as discussed above. It is observed that the amplitude of the *bc*-rotation (in red) is always smaller than that of the *ac*-rotation (in blue); $|\tau_{bc}(\theta)| < |\tau_{ac}(\theta)|$. This finding indicates, as illustrated in Eq. (2)–(3), that $|\chi_{bb} - \chi_{cc}| < |\chi_{aa} - \chi_{cc}|$. By combining the susceptibility results with this out-of-plane torque measurement, it follows that $\chi_{aa} > \chi_{bb} > \chi_{cc}$ for $T < T_N$, and $\chi_{aa} < \chi_{bb} < \chi_{cc}$ for $T > T_N$. It constitutes further evidence that the *b*-axis is the favored spin alignment direction within the plane, while the overall easy axis is aligned along the *c**-axis direction.

## *2.3 Further details of magnetic anisotropy – the nonlinear off-diagonal response and the crystalline symmetry*

An important feature revealed in Fig. 2(b) is that, for the *ac*-rotation ($\phi = 0°$; blue circles), the positive and negative amplitudes are significantly different, i.e., the crests of the sinusoidal motion of the filled-blue curve reach up to +5.3, while its troughs reach only −4.3. A similar observation is made for the open-blue curve, ranging from −18.9 to +24.7. On the other hand, for the *bc*-rotation ($\phi = 90°$; red circles), the filled-symbol and empty-symbol curves are symmetric in the positive and negative planes. To further elucidate this asymmetry, $\tau_{ac}(\theta)$ was examined by varying the field

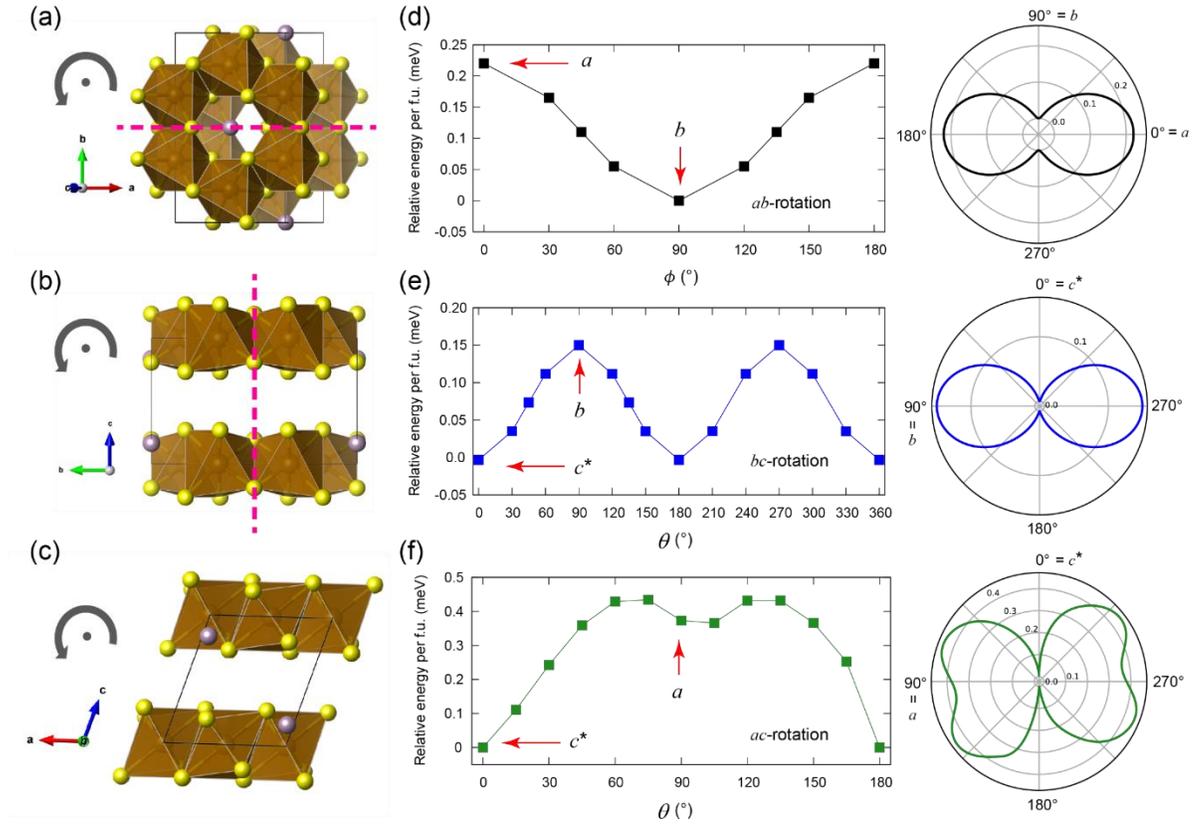

**Figure 4** (a)–(c) The crystal structure of FePS$_3$ as viewed from the direction perpendicular to the (a) *ab*, (b) *bc*, and (c) *ac* planes. The solid-black and the magenta-dashed lines indicate the unit cell and the mirror plane, respectively. The gray circular arrows show the spin rotating direction in the DFT calculation for each plane. (d)–(f) The calculated magnetic anisotropy energy profile calculated with the rotating angles of AFM spin alignment for (d) *ab*-, (e) *bc*-, and (f) *ac*-rotation. The red arrows in the left panels indicate the angles corresponding to the specific crystallographic axes. The right panels show that calculated total energies in polar coordinates. The scales of radial axes are same with the left panels. The maximum energy angle in (f) corresponds to 127°, and it is close to the direction pointing to the S ions which are a part of the octahedron cage surrounding Fe ions.

strength at $T = 50$ K. The results are presented in Fig. 3(a). For a weak field $H < 4$ T, the measured $\tau_{ac}(\theta)$ is consistent with the $\sin 2\theta$ curve (red lines). Conversely, for a strong field, $H > 4$ T, the positive values of $\tau_{ac}(\theta)$ are clearly greater than its negative values, indicating that rotating the system is easier in the range $90° < \theta < 180°$ (or, equivalently, $0° < \theta < -90°$) than in the range $0° < \theta < +90°$. The same asymmetry is also evident with the field-dependent torque, $\tau(H)$, as illustrated in Fig. 3(b), which shows that $\tau(H)$ follows the parabolic lines in positive and negative $\theta$s at low field. Under a high field, on the other hand, $\tau(H)$ significantly deviates from the parabola and appears saturated only at positive $\theta$.

We found that this observed asymmetry carries a significant physical meaning by reflecting the crystalline symmetry of this material. In the model described above, it was assumed that all the off-diagonal components of susceptibility are zero. We first noted that the asymmetries shown in Fig. 2(b) and Fig. 3(a) and (b) are indicative of a limit that violates this assumption. Thus, we presume that the asymmetry originates from the nonlinear off-diagonal components of the magnetic susceptibility. Actually, this non-zero off-diagonal contribution can be expected from such a crystal symmetry. Fig. 4(a)–(c) show the crystal structure viewed from the direction perpendicular to the (a) *ab*, (b) *bc*, and (c) *ac* planes, respectively. The magenta dashed lines represent the mirror plane. The mirror plane is well defined in the *ab*- and *bc*-projections, but not in the *ac* plane which possesses only C$_2$ rotational symmetry. Owing to the mirror symmetry plane parallel to the *ac* plane, $\chi_{ab} = \chi_{ba} = \chi_{bc} = \chi_{cb} = 0$. However, in the *ac* plane, the C$_2$ rotational symmetry without mirror symmetry can lead to non-zero off-diagonal components of susceptibility; namely, $\chi_{ac} \neq 0$ and $\chi_{ca} \neq 0$. Considering this point, Eq. (2) can be updated as follows:

$$\tau_{ac} = (\chi_{cc} - \chi_{aa})H_a H_c + \chi_{ca}H_a^2 - \chi_{ac}H_c^2, \qquad (5)$$

where $H_a$ and $H_c$ indicate the magnetic field components parallel to the *a*- and *c*-axes, respectively. This relationship is directly deduced from Eq. (1) with non-zero off-diagonal terms. If we further suppose that these off-diagonal components are nonlinear as $\chi_{ca} = \chi_{ca}^0 H_c^2$ and $\chi_{ac} = \chi_{ac}^0 H_a^2$, Eq. (5) then becomes:

$$\tau_{ac} = \frac{1}{2}(\chi_{cc} - \chi_{aa})H_0^2 \sin 2\theta + \frac{1}{4}(\chi_{ca}^0 - \chi_{ac}^0)H_0^4 \sin^2 2\theta, \qquad (6)$$

where $\chi_{ca}^0$ and $\chi_{ac}^0$ are constant. The calculated torque in Eq. (6) is plotted in Fig. 3(c) and (d) which are in good agreement with the experimental results shown in Fig. 3(a) and (b), respectively. Our experimental results together with theoretical interpretations demonstrate the capability of torque magnetometry in revealing the detailed crystalline as well as magnetic anisotropy, which can be generally applicable to other material systems. Hereby we constructed the 3-D mapping of magnetic anisotropy in terms of torque.

## *2.4 Energetics of magnetic anisotropy - DFT calculation*

Theoretically, magnetic anisotropy can also be revealed by performing total energy calculations within the relativistic non-collinear DFT scheme. Fig. 4 summarizes our results for various spin directions in which we rotate the spin alignment angles while keeping the AFM order. The computation details are given in the Methods section. Fig. 4(d)–(f) shows the magnetic anisotropy energy profiles as a function of angle changes within the *ab*, *bc*, and *ac* planes, respectively. From Fig. 4(d), it is evident that, within the *ab* plane, the spins align preferably along the *b*-axis direction rather than along the *a*-axis direction, in agreement with our torque measurements. Additionally, Fig. 4(e) and (f) show that, in the *bc* and *ac* planes, the preferred direction corresponding to the minimal relative energy is along the *c**-axis. This finding constitutes a direct and independent confirmation of the experimentally known magnetic easy axis direction [21, 24, 26, 39] in terms of energy. The energetic order of the spin alignment direction obtained via the DFT calculations is E$_{//c}$ < E$_{//b}$ < E$_{//a}$. Again, this finding is consistent with the conclusion stated above regarding the torque measurements at $T < T_N$. Therefore, the DFT calculations not only confirm the experimental results, but also reveal the quantitative energetics of the magnetic anisotropy of this material for the first time. The energy profile is depicted in polar coordinates as shown in the right panel of Fig. 4(d)–(f). A peculiar feature of these figures is the non-sinusoidal shape of the anisotropy energy observed only for the *ac*-rotation. The total energy graph shows a minimum at a rotation angle of approximately 90°. Additionally, the increasing behavior of the relative energy from $\theta = 0°$ to ~60° is not symmetric with the decreasing behavior from $\theta =$ ~120° to 180°. Thus, the magnetic anisotropy energy profile calculated on the *ac* plane exhibits only C$_2$ rotational symmetry. This mechanism also naturally explains the asymmetric torque observed in the *ac*-rotation as shown in Fig. 3.

## 3. Conclusions

The detailed magnetic anisotropy in the full 3-D space and in terms of torque and energy have been quantitatively revealed for vdW FePS$_3$. The good agreement between our results of torque measurement, model analysis, and DFT total energy calculations is impressive especially considering the small energy scale of the magnetic anisotropy energy. The observed anisotropy is a concerted result of numerous factors including the local crystal field effect, orbital occupation and shape, crystalline anisotropy, and the long-range pattern of the spin order. Our approach can be extended to other vdW magnets for the investigation of their detailed magnetic behaviors. vdW materials can host various magnetic configurations and related phenomena such as pressure-driven spin-crossover, semiconductor-to-metal transition, and superconductivity [41]. Given that the change in the magnetic field direction is the key to control magnetism, the intimate coupling of the magnetic anisotropy to the induced phase transitions will enrich the material candidates as well as their spin functionalities for all vdW material-based spintronics.

## 4. Methods

### 4.1 Single-crystal growth and characterization

Single-crystal FePS$_3$ was synthesized via chemical vapor transport. The starting materials were sealed in a quartz ampule at a pressure below $10^{-2}$ Torr. A horizontal two-zone furnace was used, with the temperatures set to 750 and 730 °C for the hot and cold zones, respectively; these temperatures were maintained for 9 days. The stoichiometry of each single crystal was confirmed using a scanning electron microscope (COXI EM30, COXEM) equipped with an energy-dispersive X-ray spectrometer (Quantax 100, Bruker). The quality and orientation of each sample were characterized using two X-ray diffractometers: a Laue diffractometer (TRY-IP-YGR, TRY SE) and a single-crystal diffractometer (XtaLAB P200, Rigaku). All the samples exhibited C2/m symmetry. The crystallographic directions were distinguished, and the axes were labeled: *a*-, *b*-, and *c\**-axis. Susceptibility was measured in a magnetic field of 1 T applied parallel to the *c\**- or *a*-axis using a commercial magnetic property measurement system (MPMS, Quantum Design). The single-crystal FePS$_3$ was mounted on a piezoresistive cantilever such that the torque generated by the magnetization was determined from the change in the resistance of the piezo material. The change in resistance was obtained using a Wheatstone bridge circuit.

### 4.2 First-principles DFT calculations

DFT calculations were performed within the generalized gradient approximation (GGA) [42] for the exchange-correlation potential. Projector-augmented wave potentials were used [43] as implemented in the VASP (Vienna Ab initio simulation package) code [44]. The wave functions were expanded with plane waves up to an energy cutoff of 450 eV, and gamma-centered 9×9×12 k-meshes were adopted. The known ground state of the zigzag-striped AFM unit cell was obtained from the experimental crystal structure [21]. To properly describe the on-site electronic correlation within Fe 3*d* orbitals, the DFT+*U* method was adopted with a charge-density-based scheme [45–50]. $U = 6.80$ eV and $J = 0.89$ eV were used [37]. To calculate the magnetic anisotropy energy, the spin–orbit interaction was taken into account within the relativistic non-collinear scheme as implemented in the VASP package. In this scheme, the wave functions and the charge densities of the zigzag-striped AFM FePS$_3$ were obtained in a collinear DFT scheme and then were used to compute the total energy as a function of the spin angle rotation. In this procedure, the size of the moment was fixed, and the collinear AFM coupling was retained to align along the given orientation.

## Acknowledgements

Y.J. was supported by the National Research Foundation of Korea (NRF) grant by the Korea government (MSIT) (Nos. NRF-2018K2A9A1A06069211 and NRF-2019R1A2C1089017). M.J.H. was supported by the National Research Foundation of Korea (NRF) grant funded by the Korea government (MSIT) (No. 2018R1A2B2005204 and NRF-2018M3D1A1058754). This research was supported by the KAIST Grand Challenge 30 Project (KC30) in 2020 funded by the Ministry of Science and ICT of Korea and KAIST (N11200128). W.K. acknowledges the support by the NRF grant (Nos. 2018R1D1A1B07050087, 2018R1A6A1A03025340). Work at the Center for Quantum Materials and SNU was supported by the Leading Researcher Program of the National Research Foundation of Korea (Grant No. 2020R1A3B2079375). A portion of this work was performed at the National High Magnetic Field Laboratory, which is supported by National Science Foundation Cooperative Agreement No. DMR-1644779* and the State of Florida.

## Author contributions

M. N., W. K. and Y. J. performed the torque measurement. D. H. K. and M. J. H. conducted the theoretical analysis and calculation. S. L., S. S. and J. G. P. synthesized the single crystals. M. J. H. and Y. J. supervised the project. M. J. H. and Y. J. wrote the manuscript together with D. H. K and all the other authors. All of the authors contributed to the data analysis and the manuscript writing.

## Conflict of interest

The authors declare that they have no conflict of interest